\documentclass[aps, superscriptaddress, longbibliography, reprint]{revtex4-2}

\pdfoutput = 1

\usepackage{graphicx}
\usepackage{bm}
\usepackage{dcolumn}
\usepackage{amssymb,amsmath}
\usepackage[colorlinks=true,allcolors=blue]{hyperref}
\usepackage{color}
\usepackage{siunitx}
\usepackage{placeins}
\usepackage{lipsum}
\newcommand{\cOut}[1]{}
\begin{document}

\title{Excitation of the Gyrotropic Mode in a Magnetic Vortex by Time-Varying Strain}

\author{Vadym Iurchuk}
\email[Corresponding author's e-mail: ]{v.iurchuk@hzdr.de}
\affiliation{Institute of Ion Beam Physics and Materials Research, Helmholtz-Zentrum Dresden-Rossendorf, 01328 Dresden, Germany}

\author{J\"urgen Lindner}
\affiliation{Institute of Ion Beam Physics and Materials Research, Helmholtz-Zentrum Dresden-Rossendorf, 01328 Dresden, Germany}

\author{J\"urgen Fassbender}
\affiliation{Institute of Ion Beam Physics and Materials Research, Helmholtz-Zentrum Dresden-Rossendorf, 01328 Dresden, Germany}
\affiliation{Institute of Solid State and Materials Physics, Technische Universit\"at Dresden, 01062 Dresden, Germany}

\author{Attila K\'akay}
\affiliation{Institute of Ion Beam Physics and Materials Research, Helmholtz-Zentrum Dresden-Rossendorf, 01328 Dresden, Germany}

\date{\today}

\begin{abstract}
We demonstrate excitation of the gyrotropic mode in a magnetostrictive vortex by time-varying strain. The vortex dynamics is driven by a time-varying voltage applied to the piezoelectric substrate and detected electrically by spin rectification at subthreshold values of rf current. When the frequency of the time-varying strain matches the gyrotropic frequency at given in-plane magnetic field, the strain-induced in-plane magnetic anisotropy leads to a resonant excitation of the gyration dynamics in a magnetic vortex. We show that nonlinear gyrotropic dynamics can be excited already for moderate amplitudes of the time-varying strain.
\end{abstract}

\maketitle

%\section{Introduction}

Magnetic vortices -- stable topological magnetic configurations -- can be spontaneously formed in confined high-symmetry magnetic micro- and nanostructures~\cite{aharoni_upper_1990, usov_magnetization_1993, shinjo_magnetic_2000}. They are considered as promising candidates for applications in next-generation spintronic memory~\cite{bohlens_current_2008}, sensor~\cite{suess_topologically_2018}, and oscillator~\cite{dussaux_large_2010, litvinenko_2020} devices.
Resonant excitation of a vortex core by in-plane rf magnetic fields results in the gyrotropic dynamics of the vortex core (VC)~\cite{guslienko_eigenfrequencies_2002, guslienkoVortexstateOscillationsSoft2005}. To reduce energy dissipation and to facilitate integration of magnetic vortices with CMOS-based components, alternative means of VC dynamics excitation are usually employed, e.g. dc-current-driven excitation by Slonczewski spin-transfer torque~\cite{slonczewskiCurrentsTorquesMetallic2002, pribiagMagneticVortexOscillator2007a} or rf-current-driven excitation via non-adiabatic Zhang-Li spin-transfer torque~\cite{zhangRolesNonequilibriumConduction2004, kasaiCurrentDrivenResonantExcitation2006}.

An interesting approach was proposed by Ostler \textit{et al.} in~\cite{ostlerStrainInducedVortex2015} to use time-varying-strain gradient to excite large-amplitude gyration dynamics in magnetostrictive microstructures with Landau-flux-closure state leading to eventual switching of the VC. This approach holds promises for considerable reduction of the energy consumption (due to the absence of currents flowing through the device) and offers a purely extrinsic means to excite the VC dynamics. However, engineering of a strain gradient, necessary for breaking the flux closure symmetry and providing an onset for the strain-driven VC dynamics, requires a sophisticated sample design, which is not straightforward to realize in real devices. In addition, when using piezoelectric materials for strain generation, large voltages are usually needed to generate sufficient strains. An alternative way of strain-driven VC dynamics by surface acoustic waves (SAW) was studied analytically and numerically by Koujok \textit{et al.} in~\cite{koujokResonantExcitationVortex2023a}.
Recent experiments reported on a local piezostrain as an efficient means to shift the VC gyrotropic frequency in a compact device with low voltages and all-electrical operation~\cite{iurchukPiezostrainLocalHandle2023}, thus providing a reliable path to strain-driven VC excitation. However, the experimental demonstration of the strain-driven excitation of the VC dynamics has not been reported so far.

In this letter, we demonstrate both experimentally and by micromagnetic simulations, excitation of a gyrotropic mode in a magnetostrictive vortex by local time-varying piezoelectric strain. The VC dynamics is excited by a time-varying voltage applied to the piezoelectric substrate and detected electrically by spin rectification measurements at subthreshold values of rf current. Micromagnetic simulations confirm that the strain-induced in-plane magnetic anisotropy in a flexed magnetic vortex leads to the VC shift from the equilibrium position. When the frequency of the time-varying strain matches the VC gyrotropic frequency for given in-plane magnetic field, the strain-induced in-plane uniaxial magnetic anisotropy (IPUA) leads to a resonant excitation of the gyration dynamics in a magnetic vortex. The dependence of the gyration frequency and VC trajectories on the in-plane bias field and the amplitude of the time-varying strain shows that nonlinear VC dynamics can be excited at moderate strain amplitudes. This result allows for an energy-efficient excitation and control of the VC gyrotropic mode in vortex-based spin-torque oscillators.

%%%%%%%%%%%%%%%%%%%%%%%%%%%%%%%%%%%%%%%%%%
%%%%%%%%%   FIG. 1
%%%%%%%%%%%%%%%%%%%%%%%%%%%%%%%%%%%%%%%%%%
\begin{figure*}[t]
\centering
    \includegraphics[width=\textwidth]{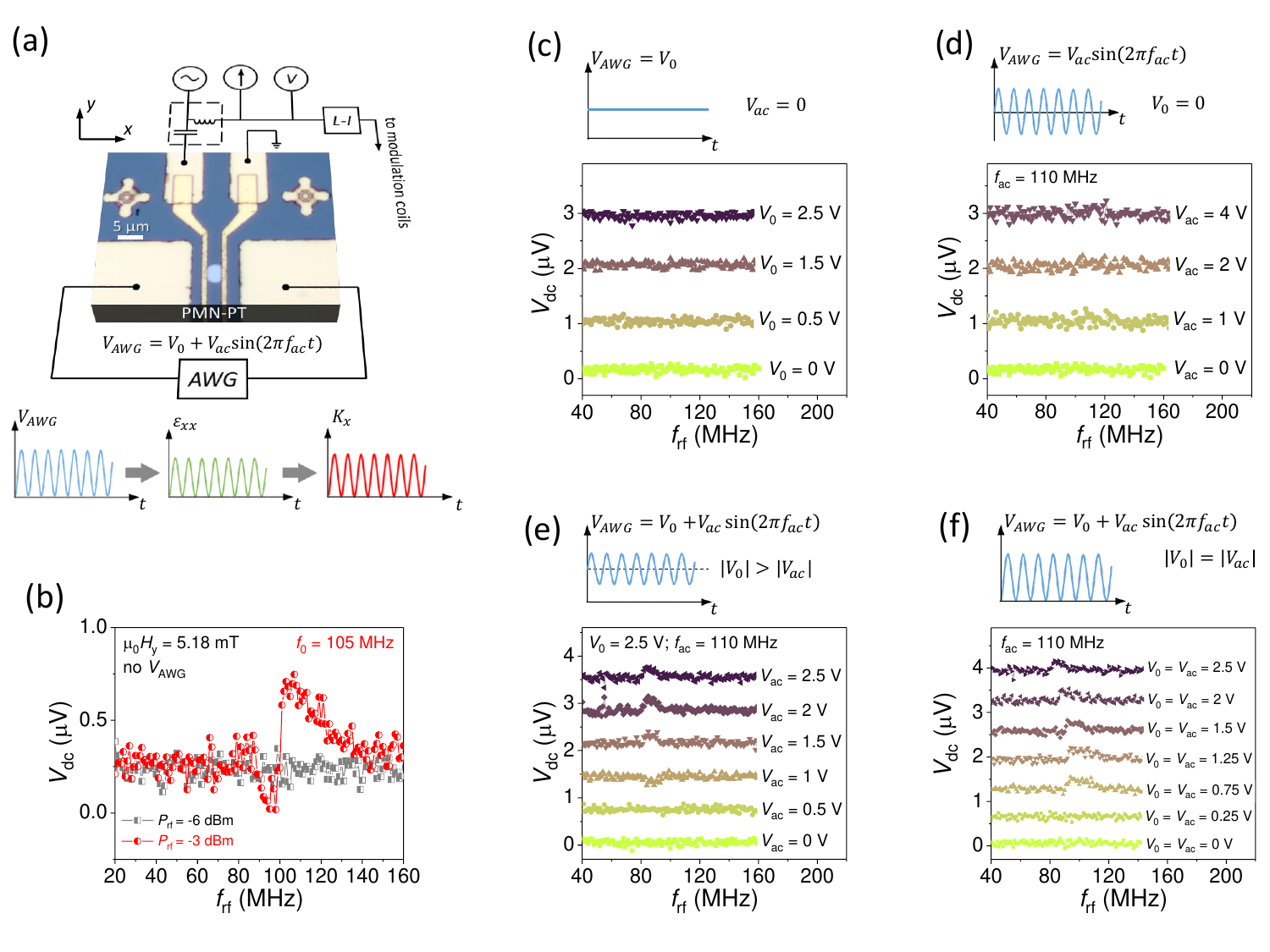}
    \caption{(a) Schematics of the experiment enabling a detection of the magnetization dynamics in a magnetostrictive vortex by a homodyne detection technique, with a simultaneous application of an time-varying voltage to the piezoelectric PMN-PT substrate. Unipolar time-varying voltage $V_{AWG}$ to PMN-PT generates the time-varying uniaxial strain $\varepsilon_{xx}(t)$, creating a time-varying magnetoelastic anisotropy $K_x(t)$ in the CoFeB disk at the same frequency $f_{ac}$. (b) Electrically detected $V_{dc}$ signal for $P_{rf}$=--3~dBm (red curve) and $P_{rf}$=--6~dBm (gray curve). (c--d) Rectified voltage $V_{dc}$ versus rf current frequency $f_{rf}$ and for different $V_{AWG}$ applied to the PMN-PT: (c) static $V_{AWG}$ with $V_{ac}$ = 0~V, $V_0$ = 0; 0.5; 1.5 and 2.5~V; (d) time-varying $V_{AWG}$ with $V_0$ = 0~V, $V_{ac}$ = 0; 1; 2 and 4~V; (e) time-varying $V_{AWG}$ with $V_0$ = 2.5~V, $V_{ac}$ = 0; 0.5; 1; 1.5, 2 and 2.5~V; (f) time-varying $V_{AWG}$ with $V_0$ = $V_{ac}$ = 0, 0.25; 0.75; 1.25; 1.5; 2 and 2.5~V.} 
    \label{fig1}
\end{figure*}

%\section{Samples and experimental setup}

The study is performed on micron-sized magnetostrictive Co$_{40}$Fe$_{40}$B$_{20}$ (hereafter CoFeB) disks grown on (011)-cut piezoelectric 0.7Pb[Mg$_{1/3}$Nb$_{2/3}$]O$_3$--0.3PbTiO$_3$ (hereafter PMN-PT) single crystals. The detailed information on the sample fabrication can be found in Ref.~\cite{iurchukPiezostrainLocalHandle2023}.
To detect the VC dynamics, we use a rf magnetotransport setup [see Fig.~\ref{fig1}(panel a)] for electrical detection of magnetization dynamics in single magnetic vortices at room temperature (see~\cite{ramasubramanian_effects_2022, iurchukPiezostrainLocalHandle2023} for more details). The standard detection technique exploits the anisotropic magnetoresistance (AMR) effect, i.e., the resistance change induced by the relative angle between the direction of the electrical current and the net magnetization of a magnetic structure~\cite{kasaiCurrentDrivenResonantExcitation2006}. An rf current injected through a bias-T into the microdisk device excites the VC gyrotropic dynamical mode, via the joint action of the spin-transfer torque and rf Oersted field, and thereby leads to a dynamical magnetoresistance oscillating at the excitation frequency. The time-averaged product of the rf current and the dynamical magnetoresistance –- which results in a rectified dc voltage $V_{dc}$ -- is measured by a conventional homodyne detection scheme using a lock-in amplifier. When the excitation frequency $f_{rf}$ matches the eigenfrequency $f_0$ of the gyrotropic mode, the resulting $V_{dc}$ is enhanced due to the dynamical magnetoresistance increase associated with the resonant expansion of the VC gyration trajectory. To improve the signal-to-noise ratio, magnetic field modulation of the dynamical magnetoresistance at the lock-in reference frequency (here 1033~Hz) was used similar to~\cite{ramasubramanian_effects_2022, ramasubramanian_thesis_2022}.

We note that in general, $V_{dc} \sim I_{rf} \Delta R$, where $I_{rf}$ is the rf current injected into the device and $\Delta R$ is the dynamical magnetoresistance change over one oscillation period~\cite{sidielvalliSizedependentEnhancementPassive2022}. The latter is directly defined by the VC trajectory opening~\cite{kasaiCurrentDrivenResonantExcitation2006}. Therefore, two ways of enhancing the $V_{dc}$ signal at resonance are possible. The first way is increase of the rf current amplitude $I_{rf}$, which also leads to the $\Delta R$ increase due to the stronger current-driven excitation of the VC. The second method is resonant enhancement of the VC gyration trajectory by external stimulus at fixed rf current $I_{rf}$. 
Here, we use time-varying strain $\varepsilon(t)$, generated by time-varying voltage applied to the piezoelectric substrate to drive the VC dynamics, which eventually leads to the VC trajectory enhancement at resonance.

In our experiment, we first characterize the rf-current-driven VC dynamics and define the threshold rf current to detect the vortex gyrotropic mode electrically. Then, we study the effect of piezostrain on the VC dynamics for subthreshold $I_{rf}$ values. We show that static strain has no effect on the rectified signal $V_{dc}$, nor has the symmetric sinusoidal time-varying strain $\varepsilon(t)$ oscillating around zero at the VC gyration frequency. On the other hand, we observe an enhanced $V_{dc}$ at the VC resonance when the symmetric sinusoidal strain is additionally biased by a static offset strain, creating a unipolar time-varying strain.

An arbitrary waveform generator (AWG) was used to apply a voltage $V_{AWG}$ with a sinusoidal time profile to the PMN-PT crystal via surface electrodes. The applied voltage is defined as $V_{AWG} = V_0 + V_{ac}sin(2 \pi f_{ac} t)$, where the first term $V_0$ is the static offset voltage and the second term is the time-varying voltage with the amplitude $V_{ac}$ and frequency $f_{ac}$. Upon application of a time-varying $V_{AWG}$, the local time-varying uniaxial strain $\varepsilon_{xx}(t)$ is generated in the PMN-PT, due to the converse piezoelectric effect. The strain, when transferred to the CoFeB disk, imposes a time-varying IPUA $K_x(t)$ at the same frequency $f_{ac}$ (see schematics at the bottom of Fig.~\ref{fig1}(a)).

To ensure that the detected VC dynamics is strain-driven, we first measure the rectified voltage as a function of the rf power for zero time-varying voltage $V_{AWG}$. Fig.~\ref{fig1}(b) shows the rectified voltage $V_{dc}$ versus the $f_{rf}$ measured for the CoFeB disk with 3.65~$\mu$m diameter at $\mu_0 H_y$ = 5.18~mT, and for two different values of the rf power $P_{rf}$ from the rf generator. For $P_{rf}$=--3~dBm, a distinctive resonance is detected at approximately 105~MHz, corresponding to the VC gyrotropic resonance frequency $f_0$. On the other hand, for $P_{rf}$=--6~dBm, no resonance is detected due to small current-driven displacement of the VC and therefore small dynamic magnetoresistance signal, which is below the noise level of the experimental setup. Keeping the rf excitation power below the detection threshold (here, $P_{rf}$=--6~dBm), we performed the measurements of the $V_{dc}(f_{rf})$ for different $V_{AWG}(t)$ applied to the PMN-PT [Fig.~\ref{fig1}(c--f)]. The insets in Fig.~\ref{fig1}(c--f) denote schematically the time profile of the voltage $V_{AWG}$ applied to the PMN-PT.

Fig.~\ref{fig1}(c) shows the rectified $V_{dc}$ spectra vs. rf current frequency measured for the same device and at the same conditions as in Fig.~\ref{fig1}(b) for different values of static $V_{AWG} = V_0$ applied to the PMN-PT. The static strain $\varepsilon_{xx}$ generated by the static voltage $V_0$ is expected to result in the $f_0$ downshift only~\cite{royInplaneAnisotropyControl2013a,iurchukPiezostrainLocalHandle2023}. Therefore, we observe no resonance for any $V_0$ value, since $P_{rf}$ is below the detection threshold. 
Fig.~\ref{fig1}(d) shows the $V_{dc} (f_{rf})$ spectra measured for $V_{AWG} = V_{ac} sin(2 \pi f_{ac} t)$ with $f_{ac}$ = 110~MHz and for different amplitudes $V_{ac}$ of the sinusoidal time-varying voltage. For this case, indeed no resonance is observed as well. However, when the time-varying sinusoidal voltage with the static offset $V_0 \geqslant V_{ac}$ is applied [see Fig.~\ref{fig1}(e,d)], 
i.e. for $V_{AWG} = V_0 + V_{ac} sin(2 \pi f_{ac} t)$, we observe a resonance peak in the $V_{dc} (f_{rf})$ spectra for increased values of $V_{AWG}$ amplitude.
One can see that we are able to detect resonantly the VC gyration only for the unipolar time-varying voltage, i.e. when the $V_{AWG}$ oscillates between zero and maximum value $V_0 + V_{ac}$ [as in Fig.~\ref{fig1}(f)] or between two positive values $V_0 - V_{ac}$ and $V_0 + V_{ac}$ [as in Fig.~\ref{fig1}(e)]. 

%%%%%%%%%%%%%%%%%%%%%%%%%%%%%%%%%%%%%%%%%%
%%%%%%%%%   FIG. 2
%%%%%%%%%%%%%%%%%%%%%%%%%%%%%%%%%%%%%%%%%%
\begin{figure*}[t]
\centering
    \includegraphics[width=\textwidth]{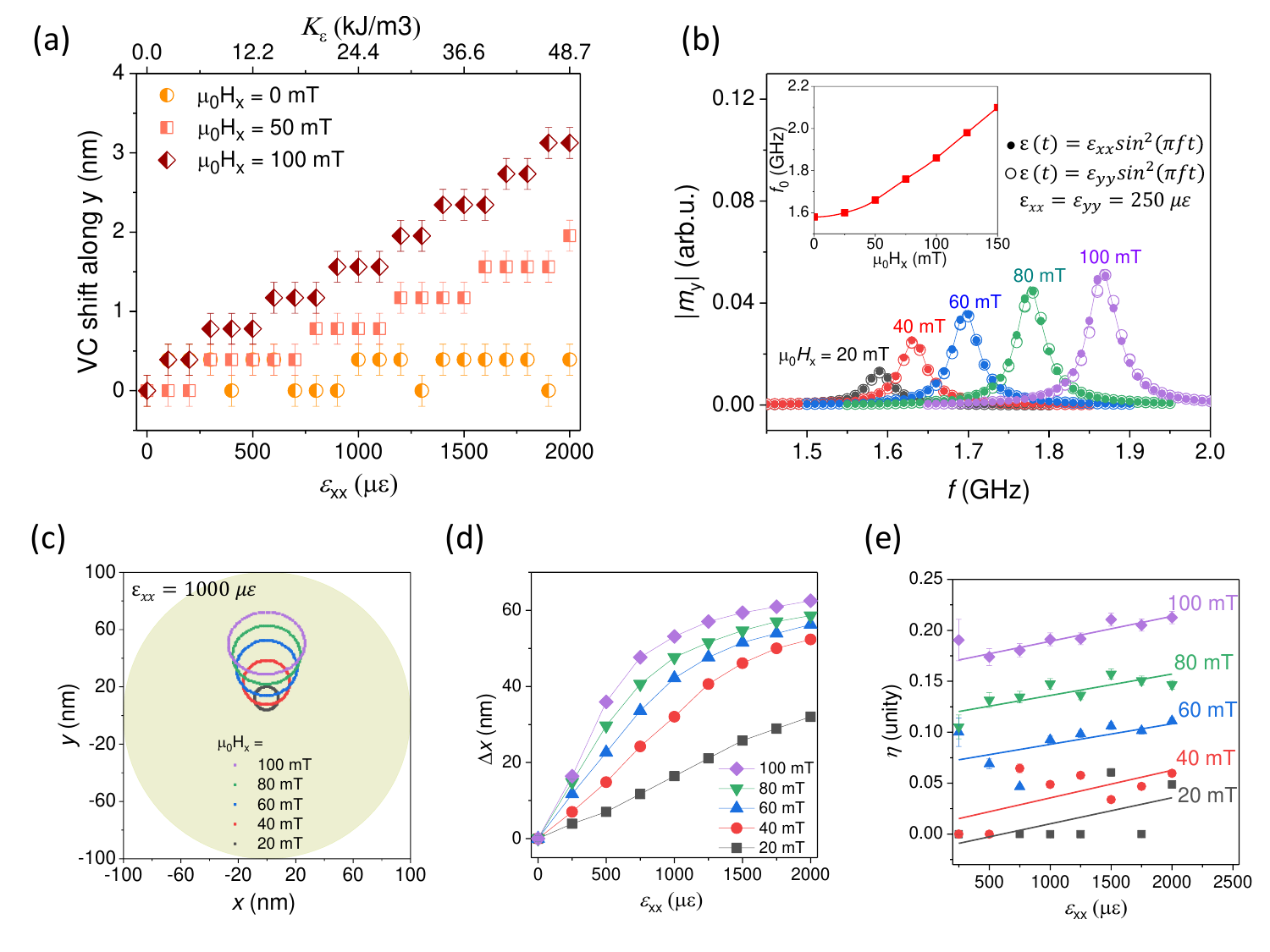}
    \caption{(a) Shift of the vortex core from the equilibrium position for $\mu_0 H_x$ = 0; 50 and 100~mT as a function of the uniaxial strain $\varepsilon_{xx}$ applied along the $x$ axis. (b) FMR-like spectra of the VC gyration dynamics excited by time-varying strain $\varepsilon(t) = \varepsilon_0 sin^2(\pi f t)$ for $\varepsilon_0 = \varepsilon_{xx}$ = 250~$\mu \varepsilon$ (solid dots) and $\varepsilon_0 = \varepsilon_{yy}$ = 250~$\mu \varepsilon$ (open circles) simulated for different values of the bias field $H_x$. The inset shows the simulated values of the VC gyration frequency $f_0$ versus $\mu_0 H_x$. (c) Simulated trajectories of the VC gyration at resonance for different $H_x$ and for $\varepsilon_{xx}$ = 1000~$\mu \varepsilon$. (d) VC trajectory opening $\Delta x$ along the $x$ axis as a function of $\varepsilon_{xx}$ for different $H_x$ values. (e) VC trajectory ellipticity $\eta$ as a function of $\varepsilon_{xx}$ for different $H_x$ values. Solid lines are linear fits.} 
    \label{fig2}
\end{figure*}

%\section{Effect of the static IPUA on the VC position}
To gain more insight into the time-varying-strain-driven gyrotropic dynamics in a magnetic vortex, and to understand why the dynamics is detectable only for unipolar time-varying voltage, we performed micromagnetic simulations of the static magnetization distribution and the magnetization dynamics in the CoFeB disks in response to static and time-varying uniaxial strain.
We use the graphics-processing-unit-accelerated \textsc{MuMax3} software~\cite{vansteenkisteDesignVerificationMuMax32014} with magnetoelastic extension~\cite{10.12688/openreseurope.13302.1}. The following parameters for CoFeB were used: saturation magnetization $M_s$ = 1700 kA/m, exchange constant $A_{ex}$ = 21 pJ/m$^3$, damping parameter $\alpha$ = 0.008, Young's modulus $Y$ = 250~GPa, saturation magnetostriction $\lambda_s$ = 65~ppm. For simplicity, we assume no magnetocristalline anisotropy in the CoFeB disk. To reduce the simulation time, we consider a much smaller CoFeB magnetic disk with radius $R$ = 100~nm and thickness $d$ = 20~nm. We use in-plane discretization into 128×128 cells for the magnetization dynamics simulations, and a finer discretization into 512×512 cells, for the computation of the static equilibrium magnetic states.
The magnetic vortex core is excited by a time-varying strain $\varepsilon(t) = \varepsilon_0 sin^2(\pi f t) \equiv \frac{\varepsilon_0}{2} + \frac{\varepsilon_0}{2} sin (2 \pi f t)$, where $f$ is the excitation frequency, and $\varepsilon$ is the uniaxial strain along $x$ ($\varepsilon_0$ = $\varepsilon_{xx}$) or $y$ ($\varepsilon_0$ = $\varepsilon_{yy}$) in-plane direction. We note that for the given $\varepsilon(t)$ time profile, the strain is oscillating at the frequency $f$ between zero and $\varepsilon_0$, i.e. during the excitation the magnetic vortex is subjected to exclusively non-negative values of strain.
Similar to~\cite{wagnerNumericalFerromagneticResonance2021, iurchukStressinducedModificationGyration2021}, the magnetization dynamics in the magnetic disk is simulated for different strain amplitudes $\varepsilon_0$ and in-plane bias fields $H_x$ over 100 gyration periods $T_0 = \frac{1}{f}$ and the final magnetization state is captured for each value of the excitation frequency $f$ in the chosen range. To study the VC trajectories as a function of excitation amplitude and bias field, the last 10 periods of time-evolution of the magnetization is recorded for a given resonance frequency of the gyrotropic mode $f_0$.

Fig.~\ref{fig2}(a) shows the $y$ coordinate of the VC position as a function of the uniaxial strain $\varepsilon_{xx}$ (and the corresponding IPUA $K_{\varepsilon} = \frac{3}{2} \lambda_s Y \varepsilon_{xx}$) for three values of external bias magnetic field applied along the $x$ axis. For $\mu_0 H_x$ = 0~mT, no VC shift is observed as expected for the symmetric vortex configuration subjected to the IPUA. Indeed, when the VC is located in the center of the disk, the effective magnetoelastic torques exerted by strain-induced IPUA on the VC are counterbalanced since IPUA acts on the equal amount of magnetic moments aligned parallel to the IPUA (here along $x$), and therefore no VC movement occurs. When the vortex symmetry is broken by an in-plane magnetic field, the IPUA leads to the shift of the VC from the equilibrium position at a given magnetic field as a result of a noncompensated net magnetoelastic torque acting on the VC. The direction of this torque coincides with the direction of the external-field torque for $\varepsilon \parallel H$ and is opposite for $\varepsilon \perp H$ (see Fig.~\ref{figS1} in the Appendix for the detailed description of an impact of IPUA on the static magnetization distribution in a magnetic vortex).
Thus the VC displacement increases with both, increasing $\varepsilon_{xx}$ and increasing $H_x$ [see Fig.~\ref{fig2}(a)]. 
The observed strain-induced VC displacement, while rather small (few nm), can nevertheless be efficiently used for the resonant excitation of the VC gyrotropic mode by time-varying strain. It is known that when the external force, which caused the VC shift from the equilibrium, is released, the VC will move towards the equilibrium on a spiral trajectory governed by the Thiele's theory~\cite{thieleSteadyStateMotionMagnetic1973, thieleApplicationsGyrocouplingVector2003}. Therefore, a resonant excitation of the gyrotropic mode in a vortex is possible by time-varying strain with an analogy to the excitation by rf field~\cite{novosadMagneticVortexResonance2005a}.

Fig.~\ref{fig2}(b) shows the ferromagnetic-resonance-absorption-like spectra of the VC dynamics excited by time-varying strain $\varepsilon(t) = \varepsilon_0 sin^2(\pi f t)$ for $\varepsilon_0 = \varepsilon_{xx}$ = 250~$\mu \varepsilon$ (solid dots) and $\varepsilon_0 = \varepsilon_{yy}$ = 250~$\mu \varepsilon$ (open circles) and for different values of the bias field $H_x$. Here and further, the strain is expressed in the dimensionless units of microstrain $\mu \varepsilon$ = $\mu$m/m~\cite{gautschiStrainSensors2002}. When the frequency of the time-varying strain $\varepsilon(t)$ approaches the eigenfrequency of the VC gyrotropic mode at a given magnetic field, a pronounced peak is observed, attributed to the resonant excitation of the VC gyration. The gyrotropic nature of the observed mode is confirmed by visualizing the VC trajectories at resonance for a given magnetic field [see Fig.~\ref{fig2}(c)]. One can see that the trajectory opening increases with increased $H_x$ i.e. when the VC is shifted closer towards the disk edge. This effect agrees with the data of Fig.~\ref{fig2}(a), where the strain-induced VC displacement for the given $\varepsilon_{xx}$ increases with increased $H_x$ due to enhanced magnetoelastic torque acting on the VC. We note that at zero magnetic field $H_x$, no gyrotropic dynamics can be excited due to the symmetry reasons described above.

Fig.~\ref{fig2}(d) shows the major axis $\Delta x$ of the VC trajectory for different values of magnetic field $H_x$ and different amplitudes $\varepsilon_{xx}$ of the time-varying strain. For the given $\varepsilon_{xx}$ range, the trajectory opening is quasilinear for $\mu_0 H_x$ = 20~mT. For $\mu_0 H_x \geqslant$ 40~mT, two strain-dependent regions can be distinguished: the linear range, and the saturation range, with the field-dependent transition between the two. We attribute the saturation range to the strain-driven large-amplitude nonlinear VC gyration dynamics. Further increase of time-varying-strain amplitude $\varepsilon_{xx}$ and/or bias magnetic field $H_x$ leads to the VC switching when reaching the critical VC velocity~\cite{lee_universal_2008}. When entering the nonlinear regime, the VC trajectory is distorted and becomes non-circular. Fig.~\ref{fig2}(e) shows the trajectory ellipticity $\eta = 1 - \frac{\Delta y}{\Delta x}$ versus amplitude of the time-varying strain for different values of $H_x$. An increase of the ellipticity is observed for increased $\varepsilon_{xx}$ at constant $H_x$ as well as for increased $H_x$ at constant $\varepsilon_{xx}$. This suggests that, for the VC dynamics driven by time-varying strain, the transition to the nonlinear dynamical range can be induced not only by simple increase of the excitation amplitude, but by shifting the VC closer to the disk edge by external magnetic fields.

%%%%%%%%%%%%%%%%%%%%%%%%%%%%%%%%%%%%%%%%%%
%%%%%%%%%   FIG. 3
%%%%%%%%%%%%%%%%%%%%%%%%%%%%%%%%%%%%%%%%%%
\begin{figure}[t]
\centering
    \includegraphics[width=0.4\textwidth]{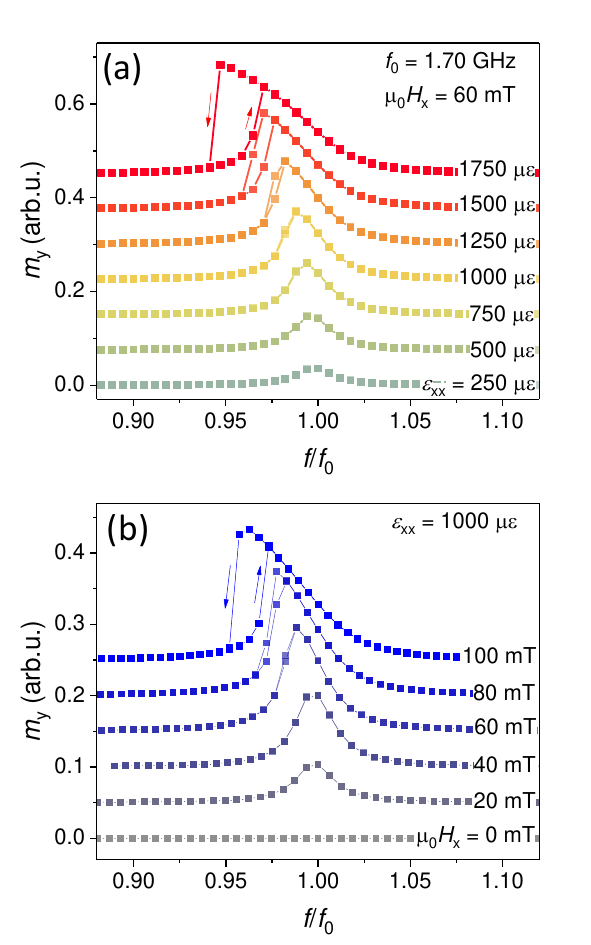}
    \caption{Frequency-swept FMR-absorption-like spectra of the VC gyration dynamics excited by time-varying strain: (a) for different values of the strain amplitude $\varepsilon_{xx}$ at fixed magnetic field $\mu_0 H_x$ = 60~mT and (b) for different values of the bias magnetic field $H_x$ at fixed $\varepsilon_{xx}$ = 1000~$\mu \varepsilon$. Arrows indicate the frequency sweep direction.} 
    \label{fig3}
\end{figure}

%\section{Strain-driven VC gyration}
Finally, we studied the effects of the bias field and the amplitude of the time-varying strain on the resonance frequency and amplitude of the VC gyration dynamics. Fig.~\ref{fig3}(a) shows the simulated absorption-like spectra for the bias field $\mu_0 H_x$ = 60~mT and for different values of strain amplitude $\varepsilon_{xx}$. For increased $\varepsilon_{xx}$ a typical transition to the nonlinear dynamics is observed, including resonance frequency shift, peak foldover and bistable behavior (hysteresis) at high excitation amplitudes~\cite{guslienkoNonlinearGyrotropicVortex2010}. Surprisingly, the same behavior is observed when the VC dynamics is excited by a fixed strain amplitude and for increased values of the bias field $H_x$ [see Fig.~\ref{fig3}(b)]. For small fields ($\mu_0 H_x \leqslant$ 40~mT), a symmetric resonance peak is observed, typical for the linear dynamics regime, whereas at higher fields ($\mu_0 H_x \geqslant$ 60~mT), the peak becomes distorted exhibiting the same signs of nonlinear dynamics as in Fig.~\ref{fig3}(a) for large $\varepsilon_{xx}$ values. As seen from the simulated spectra of Fig.~\ref{fig3} and VC trajectories of Fig.~\ref{fig2}(d,e), the critical field for the transition from linear to nonlinear dynamical regime is inversely proportional to $\varepsilon_{xx}$. As mentioned above, this result is a consequence of the increased magnetoelastic excitation of the VC, when it is shifted towards the disk edge by the bias field.

The results of the micromagnetic simulations are in good agreement with the experimental data. First, as seen in Fig.~\ref{fig1}(c), no VC resonance is observed when the VC is excited by purely sinusoidal voltage $V_{ac}sin(2 \pi f_{ac} t)$ without a dc component. Since the strain is proportional to the absolute value of the applied voltage ($\varepsilon_{xx} \sim |V_{AWG}|$), the resulting time-varying strain follows the $|sin(2 \pi f_{ac} t)|$ dependence, and therefore oscillates at twice the frequency $f_{ac}$. The simulations confirm that no gyration is excited by the time-varying strain $\varepsilon(t) = \varepsilon_{xx} sin^2 (2 \pi f t)$, where $\varepsilon_{xx}$ oscillates at 2$f$.
Second, the simulations explain the frequency shift observed experimentally for increased values of $V_{AWG}$ [see Fig.~\ref{fig1}(f)]. This frequency downshift is attributed to the excitation of the nonlinear VC gyrotropic mode. As seen in Fig.~\ref{fig3}(a), with increased amplitude of the time-varying strain, the transition to the nonlinear regime is accompanied by the peak distortion and eventual decrease of the VC gyration frequency $f_0$. Both effects are in good agreement with the experimental data of Fig.~\ref{fig1}(f).
We note that the strain driven dynamics is detected for relatively wide range of the frequencies $f_{ac}$ of the voltage $V_{AWG}$ applied to PMN-PT. We detect the resonance at approximately 90~MHz for the $f_{ac}$ range from 90 to 120~MHz. This observation suggests that the excited mode is strongly nonlinear, and therefore can be excited even by the off-resonance time-varying strain. 

In conclusion, we have demonstrated experimentally the excitation of the gyrotropic mode in a magnetostrictive vortex by time-varying piezoelectric strain. The vortex dynamics is driven by a time-varying voltage applied to the piezoelectric substrate and detected electrically by measuring the dynamical magnetoresistance of the magnetic disk at subthreshold values of rf current. Micromagnetic simulations confirm that the strain-driven excitation of the gyrotropic mode originates from the strain-induced VC shift from the equilibrium in the presence of the in-plane magnetic field, which breaks the radial symmetry of the vortex. This approach offers an extra degree of freedom for the excitation and manipulation of the magnetization dynamics of magnetic vortices for the applications in spintronic oscillators.

This study is funded by the Deutsche Forschungsgemeinschaft (DFG, German Research Foundation) within the grant IU 5/2-1 (STUNNER) – project number 501377640. Support from the Nanofabrication Facilities Rossendorf (NanoFaRo) at the IBC is gratefully acknowledged. We thank Thomas Naumann for help with the sputtering of the CoFeB thin films. We acknowledge useful discussions with Ciar\'an Fowley on the microfabrication process.

\bibliography{references}

\appendix*
\newcommand{\hbAppendixPrefix}{S}
\renewcommand{\thefigure}{\hbAppendixPrefix\arabic{figure}}
\setcounter{figure}{0}

\section{Supplementary data}
\begin{figure*}[t]
\centering
    \includegraphics[width=\textwidth]{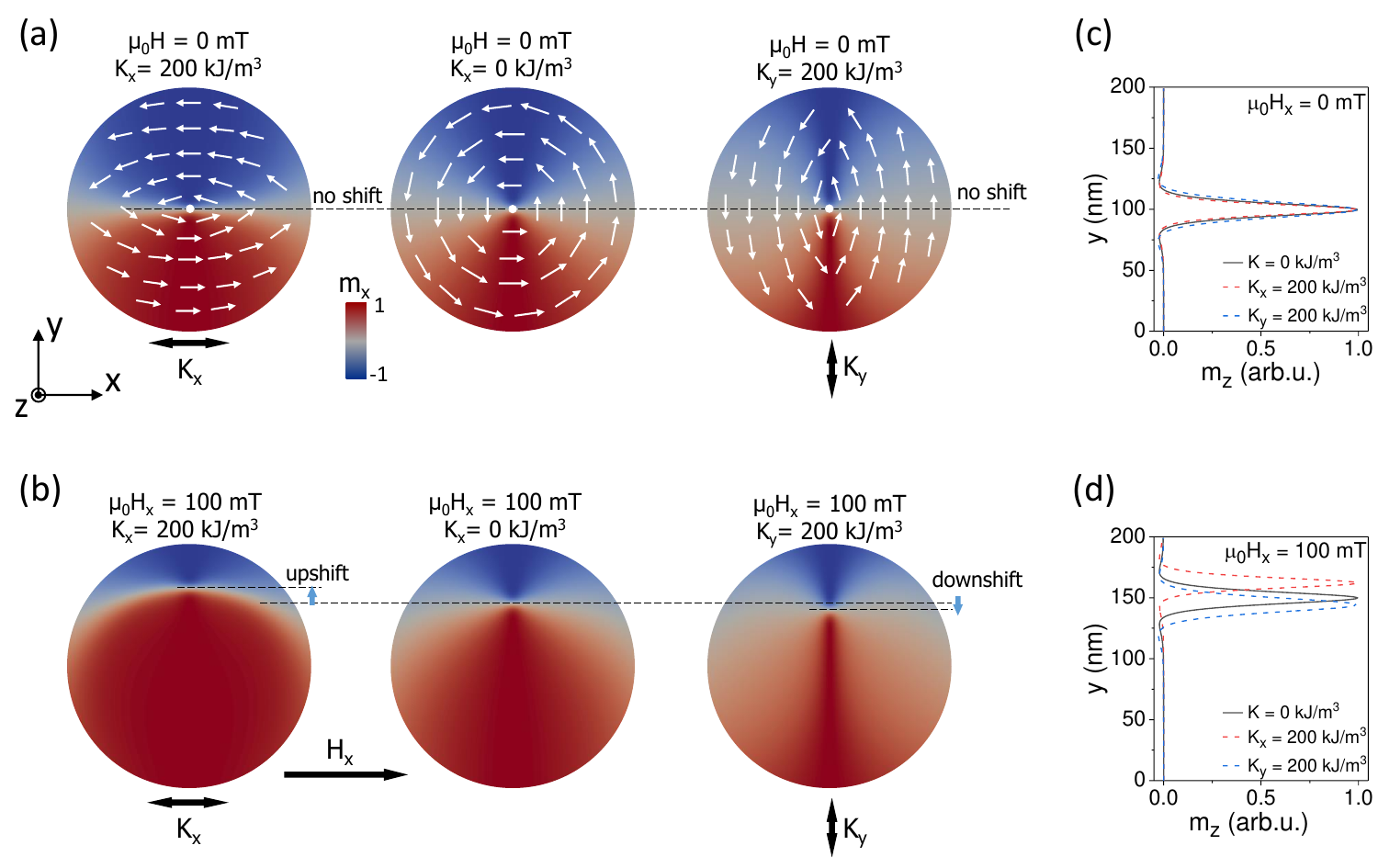}
    \caption{Effect of the in-plane uniaxial anisotropy $K$ on the magnetization distribution in a CoFeB magnetic vortex with diameter $d$ = 200~nm and thickness $t$ = 20~nm. (a) Left: $\mu_0 H$ = 0~mT; $K_x$ = 200~kJ/m$^3$ along the $x$ axis. Middle: $\mu_0 H$ = 0~mT; $K_x$ = 0~kJ/m$^3$. Right: $\mu_0 H$ = 0~mT; $K_y$ = 200~kJ/m$^3$ along the $y$ axis. White arrows denote schematically the magnetic moments within the disk. (b) Left: $\mu_0 H_x$ = 100~mT; $K_x$ = 200~kJ/m$^3$ along the $x$ axis. Middle: $\mu_0 H_x$ = 100~mT; $K_x$ = 0~kJ/m$^3$. Right: $\mu_0 H_x$ = 100~mT; $K_y$ = 200~kJ/m$^3$ along the $y$ axis. (c,d) $z$-component of the magnetization vs. $y$ coordinate for $K$ = 0~kJ/m$^3$ (solid black line); $K_x$ = 200~kJ/m$^3$ (dashed red line) and $K_y$ = 200~kJ/m$^3$ (dashed blue line) at $\mu_0 H$ = 0~mT (c) and $\mu_0 H_x$ = 100~mT (d).} 
    \label{figS1}
\end{figure*}

Fig.~\ref{figS1} shows the simulated relaxed magnetic configuration of the CoFeB disk (diameter $d$ = 200~nm; thickness $t$ = 20~nm) in the vortex state for different orientations of the in-plane bias magnetic field and in-plane uniaxial anisotropy (IPUA). Panel (a) shows the effect of the IPUA $K$ on the magnetization distribution at zero bias field. Besides the growth of the magnetic domains along the IPUA direction, no effect on the vortex core (VC) position is observed [see Fig.~\ref{figS1}(c)]. Upon application of the bias field $H$, the vortex is distorted due to the growth of the domain along the field and the corresponding shrinking of the domain with the opposite magnetization direction [see middle image in Fig.~\ref{figS1}(b)]. In such flexed magnetic vortex, the IPUA has a pronounced effect on the VC position depending on the mutual orientation of the $H$ and $K$. For $K \parallel H$ (left image in Fig.~\ref{figS1}(b)), the IPUA-driven VC shift is in the direction of the magnetic-field-induced torque, whereas for $K \perp H$ (right image in Fig.~\ref{figS1}(b)), the VC shifts in the opposite direction [see Fig.~\ref{figS1}(d)].

%%%%%%%%%%%%%%%%%%%%%%%%%%%%%%%%%%%%%%%%%%
%%%%%%%%%   FIG. S1
%%%%%%%%%%%%%%%%%%%%%%%%%%%%%%%%%%%%%%%%%%

\end{document}